\PassOptionsToPackage{dvipsnames, table, xcdraw}{xcolor}
\documentclass[table, acmsmall]{acmart}
\AtBeginDocument{%
  }

\setcopyright{none}
\renewcommand\footnotetextcopyrightpermission[1]{}
\usepackage{graphicx} 
\usepackage{supertabular,booktabs}
\usepackage{url}
\usepackage{hyperref} 
\usepackage{etc-acm}
\usepackage{mdframed}
\usepackage{color,soul}
\usepackage{listings}
\usepackage{balance}
\usepackage{array}
\usepackage{supertabular,booktabs}
\usepackage{algorithm}
\usepackage{algpseudocode}
\usepackage{amsmath}
\usepackage{multirow, multicol}
\usepackage{wrapfig}
\usepackage{makecell}
\usepackage{tabularx}
\usepackage{subcaption}

\usepackage{listing}
\usepackage{multicol}
\usepackage{color}
\usepackage{svg}

\newcommand{\exm}{\textit{extract method}}
\newcommand{\Exm}{\textit{Extract method}}
\newcommand{\llm}{large language model}
\newcommand{\llmsc}{\textsc{llm}}

\newcommand{\ppo}{\textsc{ppo}}
\newcommand{\plbart}{\textsc{plbart}}
\newcommand{\codetf}{\textsc{c}ode-\textsc{t}5}
\newcommand{\codegpt}{\textsc{c}ode\textsc{gpt}-adapt}
\newcommand{\codegen}{\textsc{c}ode\textsc{g}en}
\newcommand{\abst}{\textsc{ast}}
\newcommand{\sft}{\textsc{sft}}

\newcommand*\circled[1]{\tikz[baseline=(char.base)]{
            \node[shape=circle,draw,fill=black!75,color=black!75,text=white,inner sep=1pt] (char) {#1};}}

\begin{document}

\title{Generating refactored code accurately using reinforcement learning}

\author{Indranil Palit}
\email{indranil.palit@dal.ca}
\affiliation{%
  \institution{Dalhousie University}
  \city{Halifax}
  \state{Nova Scotia}
  \country{Canada}
}

\author{Tushar Sharma}
\email{tushar@dal.ca}
\affiliation{%
  \institution{Dalhousie University}
  \city{Halifax}
  \state{Nova Scotia}
  \country{Canada}
}

\renewcommand{\shortauthors}{Palit et al.}

\begin{abstract}
Automated source code refactoring, particularly extract method refactoring, is a crucial and frequently employed technique during software development. Despite its importance and frequent use by practitioners, current automated techniques face significant limitations. These approaches often rely on developers to identify the precise bounds of refactoring opportunities in terms of source code statements.
Also, they often do not capture the semantic context, resulting in offering no automated means to suggest meaningful method name, for instance. To address these challenges, we propose a novel reinforcement learning-based approach for fine-tuning and aligning code language models to perform automated, intelligent extract method refactoring on Java source code. Our approach fine-tunes sequence-to-sequence generative models and aligns them using the Proximal Policy Optimization (PPO) algorithm. We utilize code compilation and presence of the refactoring in the generated code as reward signals, providing a code-centric optimization process. 
Our experiments demonstrate that our approach significantly enhances the performance of large language models in code refactoring, as evidenced by both quantitative evaluation metrics such as BLEU, ROUGE, and CodeBLEU, and qualitative measures including syntactical and functional correctness. The supervised fine-tuned model, further aligned with PPO, surpasses traditional supervised fine-tuning by $11.96\%$ and $16.45\%$ in terms of BLEU and CodeBLEU scores, respectively. 
When subjected to a suite of $122$ unit tests, the number of successful tests increased from $41$ to $66$ for the reinforcement learning aligned fine-tuned Code-T5 model, highlighting the effectiveness of our approach in producing functionally correct refactorings. 

\end{abstract}


\maketitle

\section{Introduction}
Refactoring is an important software development activity that employs various techniques to enhance the structure and quality of source code without altering its functionality~\cite{Opdyke1992Refactoring, Fowler1999Refactoring}. 
By removing code smells~\cite{Fowler1999Refactoring} the practice aims to improve maintainability, encapsulating quality attributes such as readability, flexibility, and testability~\cite{ieee1990standard, chawla2015sqmma}.
Refactoring helps maintain high code quality, facilitating long-term maintainability and evolution~\cite{moser2007case}. 

\Exm{} refactoring is one of the most commonly applied refactoring techniques that involves moving a coherent code fragment from a method into a new, aptly named method~\cite{Fowler1999Refactoring}.
By creating cohesive and smaller methods, extract method refactoring not only improves code quality and maintainability but also serves as a foundation for more complex refactoring operations~\cite{Zarras2015}. 
Extract method refactoring constitutes a significant proportion, approximately $49.6$\%, of the total refactoring recommendations generated by JDeodorant~\cite{JDeodrant}, a widely recognized tool for supporting extract method operations. Furthermore, this refactoring technique has been acknowledged as a crucial operation by both open-source developers~\cite{silva2016we} and industry practitioners~\cite{van2021data}, underscoring its importance in software maintenance.

Automatically performing extract method refactoring, consist of two major steps~\cite{kramer2010legacy}. 
First, \textit{identification} of a candidate method that requires extract method refactoring; 
and second, intelligently \textit{extracting} the logic and \textit{forming} a new method with appropriate parameters, without human intervention. For the first step,
\ie{} identifying a candidate method for the refactoring,
practitioners often rely on intuition and experience.
They also utilize automated tools to assess code quality metrics and detect code smells~\cite{Sharma2018} to get aid in the decision process. The second step of automated extract method refactoring involves comprehending and extracting source code into a new method. Several approaches have been proposed to address this challenge. Hubert~\cite{hubert2019implementation} developed a method for generating extract method refactoring candidates using static code analysis tools. Maruyama~\cite{hubert2019implementation} proposed a candidate generation technique utilizing block-based slicing. Shahidi \etal{}~\cite{shahidi2022automated} introduced an algorithm for identifying, generating, and ranking extract method candidates through graph analysis. However, these approaches exhibit a few limitations. 
Specifically, most of these approaches require the developers to manually identify the bounds of a block to be refactored \ie{} start and end statements,
to perform the refactoring.
Such a reliance on human knowledge reduces the efficacy and significance of automated refactoring.
Furthermore, static analysis and metric-based methods often fail to capture latent contextual and syntactical code characteristics that could enhance the refactored code.
For instance, these approaches do not offer meaningful identifiers for the new method and its parameters.

The emergence of \llm{}s (\llmsc{}s)
has enabled the convenience in generative tasks, including code generation with high accuracy ~\cite{wang2021codet5identifierawareunifiedpretrained, ahmad2021}. The field of code generation has seen significant advancements recently, with pre-trained language models such as GitHub Copilot~\cite{githubGitHubCopilot} and Amazon Q Developer~\cite{amazonCodingAssistant} demonstrating impressive capabilities. Though such \llmsc{} perform well on many text and code generation tasks, they show mediocre performance for tasks requiring domain-specific or uncommon knowledge. 
For example, \llmsc{} have shown proficiency in generating code for known and common problems but they struggle with unfamiliar problems~\cite{chen2021evaluating}. 
Similarly, in our context,
current \llmsc{} can generate refactored code but often omit the contextual information or generate incomplete, broken, or even uncompilable code~\cite{jha2023codeattack}. Moreover, using third-party code completion services raises privacy concerns for many organizations. A notable example is Samsung Electronics~\cite{JaiVijayan_2023}, which reportedly experienced three data leakage incidents while using online code completion tools such as ChatGPT. These issues highlight the growing need for developing task-specific code generation models. 

Language models for code are sequence-to-sequence models pre-trained on large corpus of code
and can be fine-tuned for various software engineering tasks, including code summarization~\cite{ahmed2022few, sun2024source}, translation~\cite{eniser2024towards, yin2024rectifier}, completion~\cite{bairi2024codeplan, eghbali2024hallucinator}, bug localization~\cite{yang2024large}, vulnerability detection~\cite{lu2024grace, zhou2024large}, and program repair~\cite{Sharma2024}.
Despite these advancements,
to the best of our knowledge, 
the application of language models for code refactoring remains largely unexplored. 
Inspired by applying \llmsc{}s on a variety of software engineering tasks, there has been some attempts to generate refactored code using them.
For example, a recent contribution by Pomian \etal{}~\cite{Pomian2024} introduced EM-Assist, an IntelliJ IDEA plugin that leverages \llmsc{}s to generate and rank refactoring suggestions using few-shot prompting. 

Fine-tuning is another common technique to train a pre-trained model for a specific downstream task. While fine-tuning pre-trained code language models appears to be a promising solution, it has been observed that a considerable portion of programs generated by these models often fail to pass unit tests~\cite{chen2021evaluating, li2022competition, jha2023codeattack}.
Such challenges deter the adoption of the automated refactoring tools and methods.

To address these challenges, we evaluate performance of fine-tuned models and
propose a deep reinforcement learning approach that aligns fine-tuned code language models to generate refactored code by applying extract method refactoring automatically. Our approach, first, creates a dataset using state-of-the-art tools such as RefactoringMiner~\cite{Tsantalis:ICSE:2018:RefactoringMiner, Tsantalis:TSE:2020:RefactoringMiner2.0}.
We use the dataset to fine-tune four language models, pre-trained on code, using Supervised Fine Tuning (\sft{})~\cite{howard2018universal, kenton2019bert}. To enhance model performance and better align it with the objective of generating compilable code while preserving functionality, we use Proximal Policy Optimization (PPO)~\cite{schulman2017proximal} for reinforcement learning optimization.

Our reinforcement learning approach utilizes an actor-critic architecture~\cite{konda1999actor, wan2018improving}, where the actor component generates refactored code, and the critic component assesses the quality of the generated code. This architecture enables the model to learn more efficiently in the complex space of code refactoring by providing guidance on the desirability of different refactoring decisions.
The critic component incorporates discrete, non-differentiable reward signals in three stages. We first check for syntactic correctness, then assess whether the code compiles successfully, and finally, we use RefactoringMiner to detect if the desired refactoring has been applied.

To strike a balance between generating refactorings and maintaining the knowledge gained during supervised fine-tuning, we introduce a Kullback-Leibler (KL) divergence~\cite{kullback1997information, shojaee2023execution} term in the reward function. This term measures the difference between the model's current behavior and its initial behavior learned during supervised fine-tuning. By incorporating this term, we encourage the model to explore new refactoring strategies while preventing it from deviating too far from its initial understanding of code refactoring. 

Our study yielded promising results. 
The \plbart{} model,
when fine-tuned using supervised learning,
demonstrates superior performance among the chosen models
when evaluated using conventional metrics such as \bleu{},
\rouge{}, and \codebleu{}.
However, \codetf{} outperforms other models when trained with deep reinforcement learning. \textbf{We observe that combining supervised fine-tuning with deep reinforcement learning prove most effective},
compared to fine-tuning the models or training using reinforcement learning individually.
Qualitative evaluation further validates that the combination works the best, exhibiting enhanced syntactic accuracy, compilation rates, and unit test performance. 

We list the key contributions of this paper below.
\begin{itemize}
    \item We evaluate the effectiveness of supervised fine-tuned models for automatic \exm{} refactoring.
    The approach addresses the limitations of existing approaches such as manual code selection to specify the code block to-be extracted.
    \item The study presents a hybrid method that combines supervised fine-tuning with reinforcement learning optimization, specifically tailored for extract method refactoring tasks. We then evaluate the approach both quantitatively and qualitatively to ensure that it generates syntactically and semantically accurate refactorings.
    \item This study also contributes a tool for analyzing Java repositories on GitHub to create an extract method refactoring datasets with associated metadata. We provide the tool and the dataset created using it for replication and extension purposes.

\end{itemize}

\section{Background} \label{section: background}

\subsection{Supervised Fine-Tuning of Large Language Models}

Supervised fine-tuning is an add-on training for adapting pre-trained large language models (LLMs), such as CodeT5~\cite{wang2021codet5identifierawareunifiedpretrained}, to specialized tasks. 
This adaptation is achieved by training the models on domain-specific datasets, which is particularly important for enhancing their performance in tasks such as extract method refactoring. In this context, we focus on two predominant model architectures: \textit{encoder-decoder} and \textit{decoder-only} models.

Encoder-decoder models consist of two main components. 
The encoder processes the input sequence (\ie{} source code, in our case) to create a context-rich representation, which the decoder then uses to generate the output sequence (refactored code with extracted method, in our case).
This architecture is particularly useful when the input is a code snippet, and the output is the corresponding refactored version. 
The fine-tuning objective for encoder-decoder models aims to maximize the conditional probability of the correct output sequence given an input sequence. A technique called teacher forcing is employed, where the correct output token from the previous time step is fed as input to the next step. 

Decoder-only models, such as those used in GPT-like architectures~\cite{radford2019language}, operate differently. They generate each token of the output sequence directly, conditioned on all previous tokens and the input sequence, without a separate encoding phase. The training process involves presenting the combined sequence of the input code and the refactored code to the model, typically separated by a special token, \eg{} ~\textsc{[sep]}.

For both architectures, the loss function commonly used is the cross-entropy loss, calculated over the output sequence tokens. This loss function helps the model learn to predict the correct tokens in the output sequence. 

\subsection{Reinforcement Learning for Sequence Generation}

Reinforcement Learning (\rl{}) is a branch of machine learning focused on training agents to take actions in an environment to maximize some notion of cumulative reward often involving a series of decisions~\cite{shakya2023reinforcement}. 
It uses a model known as the Markov Decision Process (MDP)~\cite{puterman2014markov}, which deals with decision-making where each action is determined by steps, and outcomes are influenced by randomness. 
In \rl{}, an agent (\ie{} an autonomous entity that takes action in the given environment)
improves its decisions through trial-and-error interactions with its environment, learning from the rewards it receives based on its actions. The agent's decision-making strategy is known as the \textit{policy}, which determines the next action to take given the current situation or \textit{state}. The \textit{state} represents the current context or input on which the agent bases its decisions.

In the context of language models, \rl{} can be employed as a training mechanism. 
Here, the language model serves as the \textit{policy}, and the current text sequence is the \textit{state}. 
The model generates an action,
the next word or token,
altering the state into a new text sequence. 
The quality of the completed text sequence determines the reward, assessed either by human judgment or a trained reward model based on human preferences. Prior studies~\cite{stiennon2020learning, christiano2017deep} has shown that \textsc{SFT} serves as a reliable starting point for \rl{}. Ouyang~\etal{}~\cite{ouyang2022traininglanguagemodelsfollow} employed a similar two-stage architecture like ours and found that \rl{} performs better when initially fine tuned using \textsc{SFT}. However, none of the works focused on the applicability in the software engineering domain especially in code refactoring.

In the software engineering domain, \rl{} has been used for code completion tasks~\cite{shojaee2023execution, li2024ircoco} and code summarization~\cite{wan2018improving}. They all use actor-critic methods to train the language models for specific downstream tasks. The actor is the policy model, the main language model pre-trained or fine-tuned on code data and the critic is another component that evaluates the output generated by the actor and provides a reward signal. Based on this architecture, we formulate our problem as follows. 

In this work, we focus on aligning a fine-tuned large language model for extract method refactoring generation using Proximal Policy Optimization (\ppo{})~\cite{schulman2017proximal}---a popular actor-critic  reinforcement learning method. 
This alignment process involves several key components: the actor, the critic, rewards, the value function, and KL divergence.

The \textit{actor} in our setting is the language model itself, which generates sequences of code such as extracting methods from code snippets. It takes the current code as input and outputs a refactored version with extracted methods. The \textit{critic} is a separate component that evaluates the quality of the refactoring produced by the actor, providing a score or reward that reflects how well the generated refactoring meets the desired criteria, such as syntactically and semantically accurate refactored code.

A \textit{reward} is a numerical score assigned to each generated refactoring, indicating its quality. Higher rewards are given for refactorings that improve code properties, while lower rewards indicate poor refactoring outcomes, such as introduction of errors. These rewards guide the actor in learning to generate more desirable refactorings over time.

The \textit{value function} estimates the expected reward from a given state or step in the sequence generation process. It predicts how good the current refactoring is, considering future rewards. In practice, the value function is represented by a separate neural network head, called a \textit{value head}, which outputs a scalar value for each input state, estimating the expected cumulative reward, denoted as 
\begin{equation} \label{eq:value_func}
 V(s) = \mathbb{E} \left[ R \mid s \right], where\ R\ is the\ total\ reward
\end{equation}

\textit{Proximal Policy Optimization (\ppo{})} is the algorithm used to train the actor model. \ppo{} optimizes the model's parameters by adjusting its behavior in small, controlled steps, ensuring that changes are not too drastic. This balance between exploration (trying new refactoring strategies) and stability (maintaining effective behaviors) helps the model learn efficiently without losing its learned knowledge.

\textit{KL Divergence} (Kullback-Leibler divergence) measures the difference between the old policy $\pi_{\theta_{old}}$ and the updated policy $\pi_\theta$, ensuring that updates to the policy do not deviate excessively from the original behavior. It is calculated as:

\begin{equation} \label{eq:kl_div}
\text{KL}(\pi_{\theta_{old}} \parallel \pi_{\theta}) = \sum_{x} \pi_{\theta_{old}}(x) \log \left( \frac{\pi_{\theta_{old}}(x)}{\pi_{\theta}(x)} \right),
\end{equation}

where $\pi_{\theta_{old}}(x)$, $\pi_\theta(x)$ represent the probability distributions over the possible code refactoring actions that the model can take at a given step.

In summary, the actor generates refactoring suggestions, and the critic evaluates them using static non differential rewards that provide feedback on their quality. \ppo{} optimizes the model's behavior gradually, guided by the value function, the objective function, and loss components, while KL divergence ensures that changes remain within reasonable limits. This framework enables the fine-tuning of the language models to produce high-quality code refactorings over time.
\section{Methods}
\noindent

This section details the goal, research questions, and the approach, including setup, and metrics used to rigorously test and validate our proposed method.

\begin{figure*}
\centering
\centerline{\includegraphics[width=\textwidth]{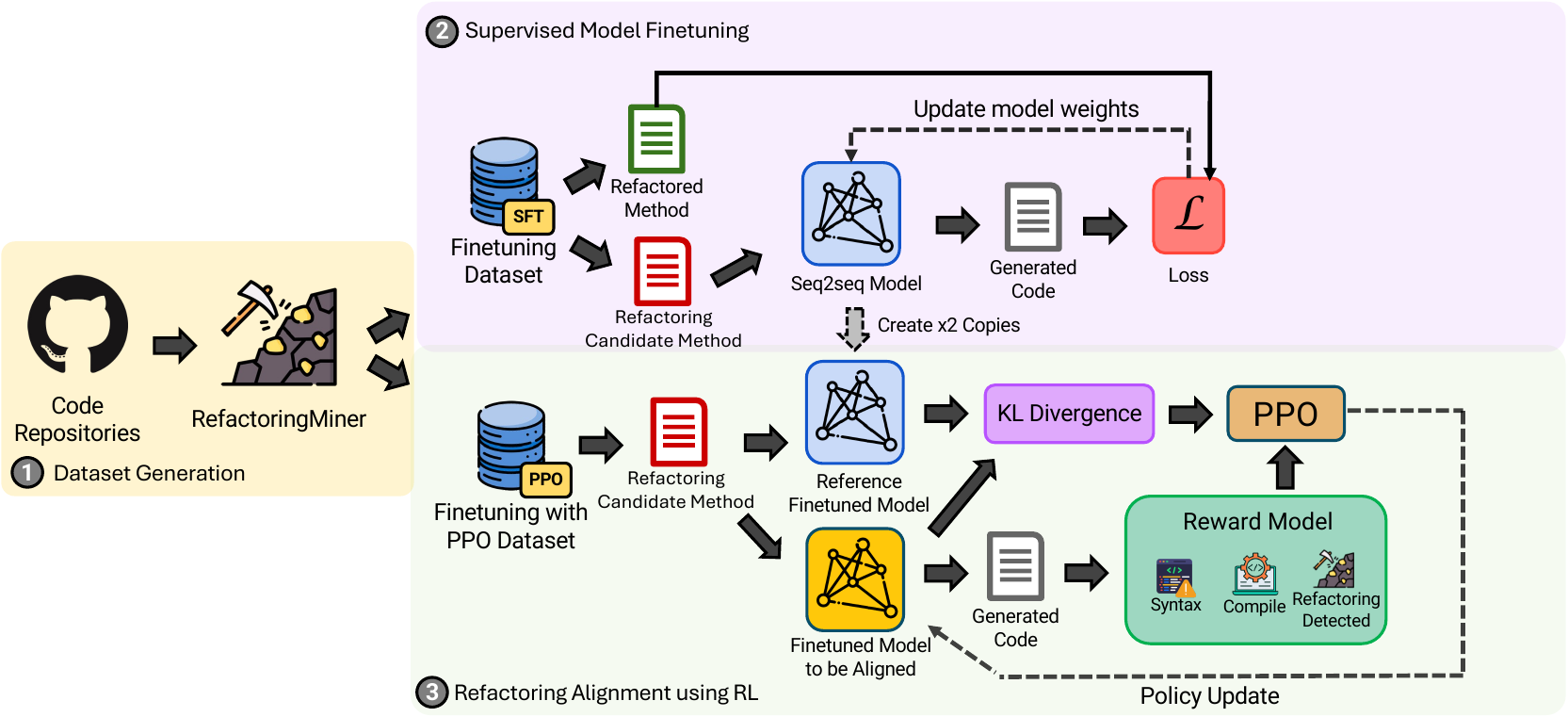}}
\caption{Overview of the proposed approach.}
\label{fig:methodology}
\end{figure*}

\subsection{Overview}

The goal of this study is to evaluate the effectiveness of fine-tuned \llmsc{}s pretrained on code and develop a deep reinforcement learning-based approach for generating code for extract method refactoring. We seek to demonstrate the effectiveness of our approach not only quantitatively but also qualitatively. We formulate the following research questions.
\begin{description}

    \item [\textbf{RQ1.}] \textit{How does supervised fine-tuning perform for extract method refactoring task?}

By answering this research question, we aim to evaluate how well does supervised fine-tuning a code \llm{} perform in automatically performing \exm{} refactoring.

    \item [\textbf{RQ2.}] \textit{How well does a reinforcement learning approach perform for automating extract method refactoring?}

This question examines whether code \llm{}s can be directly aligned using reinforcement learning techniques to effectively perform \exm{} refactoring.

    \item [\textbf{RQ3.}] \textit{How does a reinforcement learning approach, combined with fine-tuned \llm{}s, perform for automating extract method refactoring?}

This question assesses the impact of combining \ppo{} with custom reward signals on a fine-tuned model's performance in \exm{} refactoring tasks.
\end{description}

Figure~\ref{fig:methodology} illustrates our methodology. We create our dataset by using the tools such as SEART tool~\cite{seart2021data} and RefactoringMiner~\cite{Tsantalis:TSE:2020:RefactoringMiner2.0, Tsantalis:ICSE:2018:RefactoringMiner}.
Following dataset preparation, we fine-tune four different models: \codetf{} and \plbart{}, which are encoder-decoder models, and \codegpt{} and \codegen{}, which are decoder-only models. We evaluate the performance of these models using both quantitative and qualitative measures. After conducting both quantitative and qualitative evaluations, we align the pre-trained model directly using the \ppo{} algorithm. Subsequently, we align the fine-tuned model using the same \ppo{} algorithm. We systematically evaluate the applied approach using standard evaluation metrics. We also evaluate the models qualitatively using three key checks \ie{} syntactic validity, compilability, and the presence of the desired refactoring in the generated code.


\subsection{Dataset Creation} \label{section:dataset_creation}

We employ a systematic approach to identify and collect extract method refactoring instances across multiple open-source Java repositories. Step~\circled{1} in Figure~\ref{fig:methodology} shows an overview of the dataset preparation pipeline. 
We use SEART~\cite{seart2021data} tool to select a  list of repositories for analysis. 
SEART tool is a GitHub project sampling tool,
offering various commonly used filters (such as number of commits and stars).
We obtain a list of all non-forked Java repositories 
created between $2013$ and $2023$, that are active in $2024$, 
have at
least $100$ commits,
and minimum $50$ stars. 
We obtained a total of $1,618$ repositories
satisfying the criteria.

\begin{algorithm}[th!]
\caption{Procedure for Creating Dataset}
\label{algo:1}
\begin{algorithmic}[1]
    \State \textbf{Input:} List of repositories $R = \{r_1, r_2, \dots, r.
    _n\}$
    \State \textbf{Output:} JSONL file with keys ``Input'' and ``Output''
    
    \Procedure{CreateDataset}{$R$}
        \State $Data \gets \emptyset$ \Comment{Initialize the dataset as an empty set}
        \For{each repository $r_i \in R$}
            \State Retrieve branch details for $r_i$
            \State Fetch the list of commits for the given branch
            \For{each commit $c_j$ in the list of commits}
                \State Identify refactorings performed in $c_j$
                \If{extract method refactoring is detected}
                    \State Extract metadata associated with the refactoring
                    \State Extract the refactored method using the metadata
                    \State Checkout to the previous commit $c_{j-1}$
                    \State Extract the original method from $c_{j-1}$
                    \State Create output JSON object 
                    \State Append this JSON object to $Data$
                \EndIf
            \EndFor
        \EndFor
        \State Store $Data$ in a JSONL file
    \EndProcedure

\end{algorithmic}
\end{algorithm}

To iteratively process the list of repositories to prepare the dataset, we created a custom Command Line Interface (CLI) tool. 
Algorithm~\ref{algo:1} provides a pseudocode of the functionality of the tool. For each repository, we retrieve branch details and fetch the commit history. We then iterate through each commit, identifying any extract method refactorings performed using RefactoringMiner~\cite{Tsantalis:ICSE:2018:RefactoringMiner, Tsantalis:TSE:2020:RefactoringMiner2.0}. When such a refactoring is detected, the algorithm extracts relevant metadata and the refactored method from the current commit $c_j$. It then checks out the previous commit, $c_{j-1}$ to extract the original, pre-refactored method. This pair of pre- and post-refactoring methods, along with associated metadata (such as file path, class content and start and end line of the methods), is packaged into a JSON object. These JSON objects are accumulated into an array, which is ultimately stored in a JSONL file format. This approach enables the creation of a comprehensive dataset ($\mathcal{D}$) that captures the before and after states of extract method refactorings across multiple repositories.

\begin{wraptable}{ht!}{.5\columnwidth}
    \centering
    \caption{Dataset statistics} \label{tab:dataset_stats}
    \resizebox{0.5\columnwidth}{!}{
        \begin{tabular}{ccc|ccc}
            \toprule
            \multirow{2}{*}{Dataset} & \multicolumn{2}{c}{Before pre-processing} & \multicolumn{2}{c}{After pre-processing} \\
            \cmidrule(lr){2-5}
& \makecell[c]{Avg.\\ source token\\ length} &  \makecell[c]{Avg. \\ target token\\ length} &  \makecell[c]{Avg.\\ source token\\ length} &  \makecell[c]{Avg.\\ target token\\ length} & \\
            \cmidrule(lr){1-5}
\makecell[l]{$\mathcal{D}_{SFT}$}& 412.77 & 446.13 & 184.26 & 241.63\\ \addlinespace
\makecell[l]{$\mathcal{D}_{RL}$} & 410.60 & 449.09 & 187.62 & 242.13 \\ \addlinespace
            \bottomrule
        \end{tabular}    
    }
\end{wraptable}
\vspace*{\fill}

For RQ1 and RQ2, we use the entire dataset.
For RQ3,
we divide the dataset into two mutually exclusive subsets one for supervised fine tuning and the other for the aligning the fine-tuned model with deep reinforcement learning.
We divide the dataset to maintain data integrity and avoid data leak while training  for RQ3.
We divide the repository list of $1,618$ repositories, collected from the SEART tool, in half. We applied the aforementioned procedure to process both sets of repositories. This resulted in $38,441$ samples for the supervised fine tuning ($\mathcal{D}_{SFT}$) and $9,313$ samples for deep reinforcement learning ($\mathcal{D}_{RL}$).
However, the resulting datasets contained samples that exceeded the context window (maximum input sequence length) of our selected fine-tuning models.
Among these models, \texttt{Code-T5} has the smallest context window of $512$ tokens, while others support up to $2,048$ tokens. To ensure compatibility across all models, we use $512$ as our maximum context length, eliminating any samples that surpassed this $512$-token threshold. After pre processing, $\mathcal{D}_{SFT}$ contains $26,949$ samples and $6,528$ samples in $\mathcal{D}_{RL}$. 
Table~\ref{tab:dataset_stats} presents the average token length distribution for both the datasets. Finally, each of the dataset is divided in $70:20:10$ ratio for training, testing and validation.

\subsection{Training Models}

\subsubsection{Fine tuning LLMs}

We employ the following criteria to select the models for fine-tuning.
The selected models must belong to encoder-decoder or decoder-only architecture.
We exclude encoder-only models, such as CodeBERT, from our study because the encoder-only models are not well-suited for sequence-to-sequence (seq2seq) generation tasks~\cite{wang2023codet5+}. Encoder-only model architectures like \textsc{bert} are designed to understand input sequences but lack the ability to generate new ones. They're optimized for tasks like classification or feature extraction, not for producing variable-length outputs required in seq2seq tasks. Without a decoder component and autoregressive generation capability, these models can't effectively perform tasks such as translation or text generation that require producing new sequences based on input.
We select the following models, two belonging to encoder-decoder and two to decoder-only architecture family,
based on the the above-mentioned criteria.

\begin{itemize}
    \item \textbf{Code-T5}: \codetf{}~\cite{wang2021codet5identifierawareunifiedpretrained} is a pre-trained encoder-decoder model that incorporates token type information from code and employs an identifier-aware pre-training objective to better utilize identifiers. \codetf{} offers a unified framework that supports both code understanding and generation tasks, enabling multi-task learning. 
    This model has been successfully applied to various code related tasks such as code summarization~\cite{al2023extending, gu2022assemble}, code translation~\cite{kusum2022unsupervised} and vulnerability detection~\cite{paul2023, hou2024largelanguagemodelssoftware}.

    \item \textbf{PLBART}: \plbart{}~\cite{ahmad2021} 
    is a pre-trained sequence-to-sequence model that can perform a wide range of program and language understanding and generation tasks. It is trained on a large dataset of Java and Python functions along with their associated natural language text using denoising autoencoding. 
    \plbart{} has been used in various software engineering applications especially in program repair task~\cite{paul2023, wu2023effective}

    \item \textbf{CodeGPT-adapt}: \codegpt{}~\cite{lu2021codexgluemachinelearningbenchmark} is a GPT-2-based decoder-only Transformer model for code completion, pre-trained on Python and Java code from CodeSearchNet datasets. It learns code structure and syntax through pre-training, enabling it to generate code automatically.
    It has been widely used for code generation tasks such as code completion~\cite{li2024ircoco, hou2024largelanguagemodelssoftware}.

    \item \textbf{CodeGen}: \codegen{}~\cite{nijkamp2022codegen} is a Transformer-based autoregressive language model trained on natural language and programming language datasets. It employs next-token prediction as its learning objective and has shown outstanding performance in program synthesis tasks~\cite{christopoulou2022pangu}.

\end{itemize}

Step~\circled{2} in Figure~\ref{fig:methodology} illustrates the \textsc{sft} strategy employed for extract method refactoring. To train the encoder-decoder models, the pre-refactored code is first tokenized to serve as the input sequence. After a forward pass through the model, output tokens are generated and decoded using the same tokenizer. The resulting method is then compared to the ground truth, which includes both the extracted method and the modified original method post-refactoring. The model weights are updated based on the cross-entropy loss computed between the predicted and ground truth methods.

For decoder-only models, the training process is similar, with the key difference being in the format of the input. In this case, the input sequence is formed by concatenating the pre-refactored code and the ground truth output, separated by a special \textsc{[SEP]} token. This format enables the model to learn from both the context of the original code and the desired output sequence in a single input representation.

\subsubsection{Aligning the models with RL}

In this study, we fine-tune and align the selected \llm{}s for \exm{} refactoring using \rl{} techniques (step~\circled{3}). 
We model the code transformation problem as a Markov Decision Process (MDP).
We define the \textit{state} as the set of all possible code representations and the state transition function as appending the chosen refactored token to the current sequence.

Algorithm~\ref{algo:2} describes the pseudocode for aligning the fine tuned language model for extract method refactoring task. The algorithm starts with an initial policy (decision-making strategy) and a value function (which estimates how good a particular state is). It then goes through multiple training iterations to improve these over time. In each iteration, we sample a batch of code snippets from our \rl{} dataset. For each snippet, \textit{i.e.} the pre-refactored method, we use the current policy to generate a sequence of refactoring actions. To assess the quality of the sequence generated at each training step, we compute a reward based on three factors: syntactic correctness, compilation success, and whether the action is recognized as a valid refactoring. 

The reward function plays a crucial role in evaluating the quality and correctness of the refactoring suggestions produced by the model.
Our reward function consists of three key components, each addressing a specific aspect of the refactoring process:

\begin{enumerate}
    \item \textit{Syntactic Correctness:} We assess the presence of errors in the refactored code.
    For this purpose, we check the presence of error nodes in the Abstract Syntax Tree (\abst{}) generated by \textit{tree-sitter} of the generated code.
    \begin{equation} \label{eq:syntax}
        R_{syntax} = 
        \begin{cases}
            +1 & \text{if no error nodes} \\
            -1 & \text{if error nodes present}
        \end{cases}
    \end{equation}

    \item \textit{Compilation Success:} We verify whether the refactored code compiles successfully. While the compiler automatically checks for syntactic issues, separating syntactic correctness from compilation success allows us to provide the \rl{} model with more granular feedback. This distinction is important because refactored code might be syntactically correct but still fail to compile due to semantic errors.

    \begin{equation} \label{eq:compile}
        R_{compile} = 
        \begin{cases}
            +1 & \text{if code compiles} \\
            0 & \text{if code fails to compile}
        \end{cases}
    \end{equation}

    \item \textit{Refactoring Detection:} We validate the presence of extract method refactoring in the generated code using RefactoringMiner.
    \begin{equation} \label{eq: detect}
        R_{detect} = 
        \begin{cases}
            +1 & \text{if detected by RefactoringMiner} \\
            -1 & \text{if not detected}
        \end{cases}
    \end{equation}
\end{enumerate}

The sum of these individual components gives us the total reward for a given refactoring suggestion.

\begin{equation}
    R_{total} = R_{syntax} + R_{compile} + R_{detect}
\end{equation}

This reward function encourages the language model to generate syntactically correct, compilable code that successfully implements the extract method refactoring. 

The value head is used to estimate the value of the current state using the value function as shown in Equation~\ref{eq:value_func}. The algorithm then calculates how much better or worse each action was than expected (the \textit{advantage}). This information is used to update the policy, aiming to increase the probability of actions that led to high rewards. However, to ensure stable learning, the algorithm checks how much the new policy differs from the old one using a measure called KL divergence as described in equation~\ref{eq:kl_div}. If the difference is too large, the update is adjusted to prevent drastic changes. Finally, the value function is updated to better predict future rewards. By repeating this process many times, the algorithm gradually improves its ability to make good refactoring decisions. 

\begin{algorithm}
\caption{DRL Training for Extract Method Refactoring with KL Divergence}
\label{algo:2}
\begin{algorithmic}[1]
\Require Initial policy $\pi_\theta$, value function $V_\phi$, KL divergence coefficient $\beta$, weights $w_1, w_2, w_3$
\For{each training iteration}
\State Sample batch of code snippets from dataset
\For{each code snippet $x$}
\State Generate a refactored code snippet using current policy $\pi_\theta$
\State Compute syntactic correctness: $R_{syntax}$ using Eq.~\ref{eq:syntax}
\State Compute compilation success: $R_{compile}$ using Eq.~\ref{eq:compile}
\State Compute refactoring validity: $R_{detect}$ using Eq.~\ref{eq: detect}

\State Compute total reward: $R_{total} = w_1 \cdot R_{syntax} + w_2 \cdot R_{compile} + w_3 \cdot R_{detect}$
\State Estimate the value of the current state: $V_\phi(s)$
\State Calculate advantage: $A = R_{total} - V_\phi(s)$
\EndFor
\State Compute policy update to maximize:
\State \quad $J(\theta) = \mathbb{E}\left[\frac{\pi_{\theta}(a|s)}{\pi_{\theta_{old}}(a|s)} A\right] - \beta \cdot \text{KL}(\pi_{\theta_{old}} || \pi_{\theta})$
\State Apply the update to the policy: $\theta_{old} \leftarrow \theta$
\State Update value function to minimize: $L(\phi) = \sum (R_{total} - V_\phi(s))^2$
\EndFor
\end{algorithmic}
\end{algorithm}

\subsubsection{Fine tuning setup.} 

We fine-tune the supervised model for $10$ epochs on the dataset $\mathcal{D}$ for \textit{RQ1}, and on the dataset $\mathcal{D}_{SFT}$ 
for \textit{RQ3}. The training is conducted with a global batch size of $16$, using the Adam optimizer~\cite{kingma2017} with an initial learning rate of $1.33 \times 10^{-5}$. For aligning the models with \rl{}, we utilize and extend the \textsc{trl} Python library, which is widely used for training transformer language models with reinforcement learning. The generation parameters are set with \textit{min\_tokens} as $-1$ and \textit{max\_tokens} as $512$. The training consists of $20,000$ steps, with the model undergoing $10$ \ppo{} optimization epochs for each step. A global batch size of $16$ is maintained, and the Adam optimizer~\cite{kingma2017} is employed. We apply Adaptive KL control with an initial KL coefficient of $0.2$. All experiments are conducted with a fixed seed value to ensure reproducibility and are performed on nodes of a High Performance Computing (HPC) cluster, utilizing $2$ V100-32GB GPUs and $32$ GB of RAM.

\subsection{Evaluation}
In this section, we summarize the metrics commonly used for code generation tasks.
Also, we provide details about qualitative evaluation that goes beyond the standard metrics.

\subsubsection{Evaluation Metrics}

To assess the effectiveness of our models quantitatively, we utilize established metrics from natural language processing field \bleu{}~\cite{papineni2002bleu} and \rouge{}~\cite{lin2004rouge}, as well as specialized metrics tailored for code evaluation, \codebleu{}~\cite{ren2020codebleumethodautomaticevaluation} and syntax match score~\cite{zhu2022xlcostbenchmarkdatasetcrosslingual}. 
The widespread adoption of these metrics in academic research for evaluating generative models supports our decision to use them.

\subsubsection{Qualitative Evaluation} \label{subsubsection:qualitative}

To evaluate the effectiveness of our fine-tuned code language models in performing \exm{} refactoring, we construct a diverse test suite encompassing various 
complexity levels to ensure a thorough evaluation of the model's refactoring capabilities. To create the test cases for evaluation, we first identified pre-refactored original methods from the test sets of each of the dataset as mentioned in Section~\ref{section:dataset_creation}. We then selected 150 methods at random from $4,001$ test split samples ($2,695$ for \textsc{sft} and $1,306$ for \rl{}). Among these methods, few were very trivial like one or two liners and we discarded such methods. Trivial cases were removed because they do not effectively test the model's ability to handle complex refactoring tasks, providing limited insight into its true capabilities. Finally, we collected $122$ such methods which underwent extract method refactoring across various repositories. 

A significant challenge in creating unit test cases is the lack of corresponding unit tests for many methods in the selected repositories.
To address this, we leveraged gpt-$4$o (version: gpt-4o-2024-05-13) API~\footnote{https://platform.openai.com/docs/models/gpt-4o}
to generate unit tests and corresponding data. This approach aligns with recent research demonstrating the promising results of using language models for test case generation~\cite{tufano2020unit, nashid2023retrieval}. We specifically employed the \texttt{ChatTester} framework proposed by Yuan \etal{}~\cite{yuan2024chatgpt}, to generate unit tests for our samples. The framework utilized the class context of the smelly method, extracted as per Algorithm \ref{algo:1}, to create relevant unit test cases. The authors manually validated these generated test cases to ensure their quality and relevance. 
All the qualitative samples and corresponding test cases can be found in our replication package.

This combination of qualitative testing and quantitative analysis provides a systematic and objective assessment of our model's performance in extract method refactoring tasks. The multi-faceted evaluation approach allows for a comprehensive understanding of the model's capabilities and limitations across various \exm{} refactoring scenarios.

\section{Results}

\noindent
\subsection{RQ1: How does supervised fine-tuning perform for extract method refactoring task?}

The research question aims to evaluate the performance of the fine-tuned models
for the refactored code generation.
The first part of Table~\ref{tab:all_metrics}
(\ie{} \textsc{pt + sft} column)
presents the results obtained by the considered models for the refactored code generation task. 
The results presented in the table demonstrate that the \plbart{} model outperforms other models metrics and code-specific evaluation measures. Specifically, \plbart{} achieves the highest scores on the \bleu{} and \rouge{} metrics, which assess the lexical and semantic similarity of the generated text to the ground truth. Crucially, \plbart{} also exhibits superior performance on the \codebleu{} metric, which captures the syntactic and structural fidelity of the generated code snippets.

\begin{table}[ht!]
    \centering
    \caption{Experimental results for different learning objectives. Here, PT, SFT, and RL refer to pre-trained, supervised fine-tuned, and reinforcement learning-based models}
    \resizebox{\columnwidth}{!}{
        \begin{tabular}{cccc|ccc|ccc}
        \toprule
        \multirow{2}{*}{Models} & \multicolumn{3}{c}{PT + SFT (RQ1)} & \multicolumn{3}{c}{PT + RL (RQ2)} & \multicolumn{3}{c}{PT + SFT + RL (RQ3)} \\
        \cmidrule(lr){2-4}
        \cmidrule(lr){5-7}
        \cmidrule(lr){8-10}
        & \makecell[c]{BLEU} & \makecell[c]{ROUGE} &\makecell[c]{CodeBLEU} & \makecell[c]{BLEU} & \makecell[c]{ROUGE} &\makecell[c]{CodeBLEU} & \makecell[c]{BLEU} & \makecell[c]{ROUGE} &\makecell[c]{CodeBLEU} \\
        \midrule
        \makecell[l]{Code-T5} & \makecell[c]{$67.80$} & \makecell[c]{$77.49$} &\makecell[c]{$53.13$} & \makecell[c]{$38.80$} & \makecell[c]{$37.62$} & \makecell[c]{$31.99$} & \makecell[c]{$\textbf{$75.91$\textsuperscript{$\bigstar$}}$} & \makecell[c]{$\textbf{$79.92$\textsuperscript{$\bigstar$}}$} & \makecell[c]{$\textbf{$61.87$\textsuperscript{$\bigstar$}}$} \\ \addlinespace
        \makecell[l]{PLBART} & \makecell[c]{$\textbf{68.28}$} & \makecell[c]{$\textbf{80.62}$} &\makecell[c]{$\textbf{55.66}$} & \makecell[c]{$30.21$} & \makecell[c]{$29.56$} & \makecell[c]{$22.48$} & \makecell[c]{$71.20$} & \makecell[c]{$69.68$} & \makecell[c]{$58.17$} \\ \addlinespace
        \makecell[l]{CodeGPT-adapt} & \makecell[c]{$62.68$} & \makecell[c]{$65.76$} &\makecell[c]{$49.29$} & \makecell[c]{$27.68$} & \makecell[c]{$30.76$} & \makecell[c]{$20.29$} & \makecell[c]{$64.96$} & \makecell[c]{$67.82$} & \makecell[c]{$47.99$} \\ \addlinespace        
        \makecell[l]{CodeGen} & \makecell[c]{$59.32$} & \makecell[c]{$63.74$} &\makecell[c]{$42.11$} & \makecell[c]{$34.32$} & \makecell[c]{$33.74$} & \makecell[c]{$27.11$} & \makecell[c]{$\textbf{61.59}$} & \makecell[c]{$\textbf{60.52}$} & \makecell[c]{$\textbf{46.67}$} \\
        \bottomrule
        
        \end{tabular}
    }
    
    \label{tab:all_metrics}
\end{table}

Furthermore, a comparative analysis reveals that \plbart{} substantially outperforms the \codetf{} model, achieving a $4.54\%$ higher \codebleu{} score. The performance gap is even more pronounced when contrasted with the \codegpt{} model, for which \plbart{} demonstrates an $12.98\%$ improvement on the \codebleu{} metric. These findings suggest that the \plbart{} model was successful in generating extract method refactored outputs that closely resemble the ground truth in terms of syntactic correctness.

To gain a more thorough understanding of the validity and robustness of the generated refactored outputs, additional validation checks and analyses are necessary. As detailed in Section~\ref{subsubsection:qualitative}, we
create a manually validated dataset for qualitative evaluation.
The dataset contains $122$ samples with before and after refactored code and corresponding test cases.
We use this qualitative dataset to check whether the trained model generates code without any syntactic and compilation errors, whether the generated code has extract method refactoring, and to what extent the generated code is passing the test cases.
Table~\ref{tab:exresults} presents the obtained results. 
For RQ1, notably, fine-tuned \codetf{} achieved the highest performance in qualitative evaluation. This supports our assertion that relying solely on quantitative metrics may yield misleading results and potentially produce low-quality refactored code.

\begin{boxH}
\textbf{RQ1 Summary:}
Fine tuning code \llm{}s show  an effective way to teach language models to generate refactored code automatically. Specifically, \plbart{} outperform other models in all the considered metrics. It show significant improvements over \codetf{} and \codegpt{}, particularly in \codebleu{} scores. However, qualitative evaluation reveals that \codetf{} performs best in generating syntactically correct and functionally valid refactored code. 

\end{boxH}

\subsection{RQ2: How well does a reinforcement learning approach perform for automating extract method refactoring?}

This research question aims to evaluate the application of \rl{} on the refactoring task when applied on pre-trained \llmsc{}.
The second part of the Table~\ref{tab:all_metrics} (\ie{} PT + RL column) shows
the obtained results.
Our results show that \codetf{} model demonstrates superior performance across all evaluation metrics compared to other language models when trained using \rl{}.

Interestingly, unlike the results observed in RQ1 with traditional fine-tuning, direct fine-tuning using \rl{} with Proximal Policy Optimization (\ppo{}) does not perform well. This outcome may be attributed to the complexity of the \exm{} refactoring task and the potential mismatch between the \rl{} objective and the nuanced requirements of code refactoring. Fine-tuning language models that have been pre-trained on tasks other than code refactoring directly using non-differentiable rewards poses challenges. The disparity between the pre-training task and the target task of code refactoring makes it difficult to effectively train the models using \rl{} techniques. 

\begin{table}[ht!]
\centering
\caption{Qualitative evaluation of fine tuned models}
\label{tab:exresults}
\rowcolors{2}{gray!25}{white}
\begin{tabular}{p{0.5cm}p{3.75cm}|%
>{\raggedleft\arraybackslash}p{1.9cm}%
>{\raggedleft\arraybackslash}p{1.9cm}%
>{\raggedleft\arraybackslash}p{1.9cm}%
>{\raggedleft\arraybackslash}p{1.9cm}%
>{\raggedleft\arraybackslash}p{2.5cm}%
}
&\textbf{Models} & \textbf{Syntactically correct (\%)}  & \textbf{Refactoring detected (\%)} & \textbf{Compile successfully (\%)} & \textbf{\# of unit tests passed (out of $122$)} \\ \midrule
&Code-T5 (FT) & $ \textbf{78.6} $  & $\textbf{66.4}$ & $\textbf{72.1}$ & $\textbf{41}$ \\
&PLBART (FT) & $76.9$ & $63.8$ & $69.5$ & $38$ \\
&CodeGPT-adapt (FT) & ${77.5}$ & ${64.3}$ & ${70.1}$ & $39$ \\
\multirow{-4}{*}{RQ1}&CodeGen (FT) & $78.3$ & ${65.1}$ & $71.2$ & $40$ \\ \midrule
&Code-T5 + RL & {21.4}  & {20.2} & {22.3} & {21} \\
&PLBART + RL & ${21.7}$ & ${18.9}$ & ${21.5}$ & ${16}$ \\
&CodeGPT-adapt + RL& ${19.7}$ & ${14.1}$ & ${20.2}$ & $14$ \\
\multirow{-4}{*}{RQ2}&CodeGen + RL & $23.6$ & ${19.2}$ & $21.1$ & $9$ \\ 
\midrule
&Code-T5 (FT) + RL & $\mathbf{85.7}$  & $\mathbf{74.9}$ & $\mathbf{79.8}$ & $\mathbf{66}$ \\
&PLBART  (FT) + RL & ${82.4}$ & ${71.6}$ & ${76.3}$ & ${58}$ \\
&CodeGPT-adapt  (FT) + RL& ${83.1}$ & ${72.2}$ & ${77.5}$ & $61$ \\
\multirow{-4}{*}{RQ3}&CodeGen  (FT) + RL & $84.3$ & ${73.5}$ & $78.6$ & $63$ \\ \bottomrule
\end{tabular}
\end{table}
 
Qualitatively also, as shown in \textit{RQ2} of Table~\ref{tab:exresults}, the generated refactorings exhibit poor quality. The \rl{} method's poor performance in functional areas highlights a misalignment with the refactored code's true requirements. This suggests that the \rl{} reward signals may insufficiently penalize syntactic and semantic errors, resulting in models to produce functionally valid code. A potential explanation for this behavior could be that the \rl{}, performed on a generic pretrained language model, may not receive appropriate reward signals from our reward framework or the non-score rewards (KL divergence penalty). This hypothesis can be corroborated by examining Figure~\ref{fig:rl_stats}. Figure~\ref{fig:std_reward} illustrates the persistent high standard deviation of rewards for the \rl{} fine-tuned model throughout increasing training steps. This trend indicates that the reward signals fail to effectively steer the model towards optimal performance. Concurrently, Figure~\ref{fig:kl_penalty} reveals an upward trajectory in the KL-Divergence penalty over time. This escalation suggests a growing divergence between the trained model and the reference model, further supporting our hypothesis that the current reward system may be inadequate for guiding the model towards generating functionally sound code refactorings. 

\begin{boxH}
\textbf{RQ2 Summary:} Generating refactored code from a pre-trained model directly aligned with \rl{} does not produce comparable results to the corresponding fine-tuned models as shown in RQ1 quantitatively or qualitatively.
\end{boxH}

\subsection{RQ3: How does a reinforcement learning approach, combined with fine-tuned \llm{}s, perform for automating extract method refactoring?} 

In this research question, we aim to evaluate the efficacy of applying \rl{} to generate refactored code, focusing on model performance when fine-tuned using a combination of supervised fine-tuning (\textsc{sft}) and \rl{} objectives.
Specifically, we start with the trained fine-tuned models from RQ1, and train them with \ppo{} and reward from a feedback system to observe any improvements in the models compared to their fine-tuned counterparts.

Table~\ref{tab:all_metrics} presents the results in the column titled PT + SFT + RL
along with results obtained in other settings as discussed in RQ1 and RQ2. The results demonstrate that \textbf{the most effective outcomes are achieved when models are trained using both \textsc{sft} and \rl{} objectives}. This combined approach lead to significant improvements across various metrics. Specifically, we observed an approximate 10\% increase in \codebleu{} compared to models trained solely with \textsc{sft}, and an 11\% improvement over those trained exclusively with \rl{}. Similar performance gains were noted in other metrics, including \bleu{} and \rouge{}.
The superiority of the combined approach can be attributed to the complementary nature of \textsc{sft} and \textsc{rl}. \textsc{sft} excels at identifying inherent patterns and structures within data, primarily utilizing large labeled datasets. In contrast, \textsc{rl} adapts through environmental interactions, optimizing predefined reward metrics. By integrating these methodologies, our model can effectively navigate dynamic contexts while capturing underlying data patterns.

Our combined approach demonstrated a significant improvement in the evaluated quality metrics. The number of successfully passing test cases increased substantially, rising from $41$ in the best-performing model, \codetf{}, to $66$---a significant improvement of approximately $61\%$. Additionally, RefactoringMiner identified an increased number of cases, from $87$ to $98$. These results highlight the efficacy of \rl{} in producing accurate extract method refactored code.

\begin{figure*}[ht!]
    \centering
    \begin{subfigure}[t]{0.45\textwidth}
        \centering
        \includegraphics[height=1.5in]{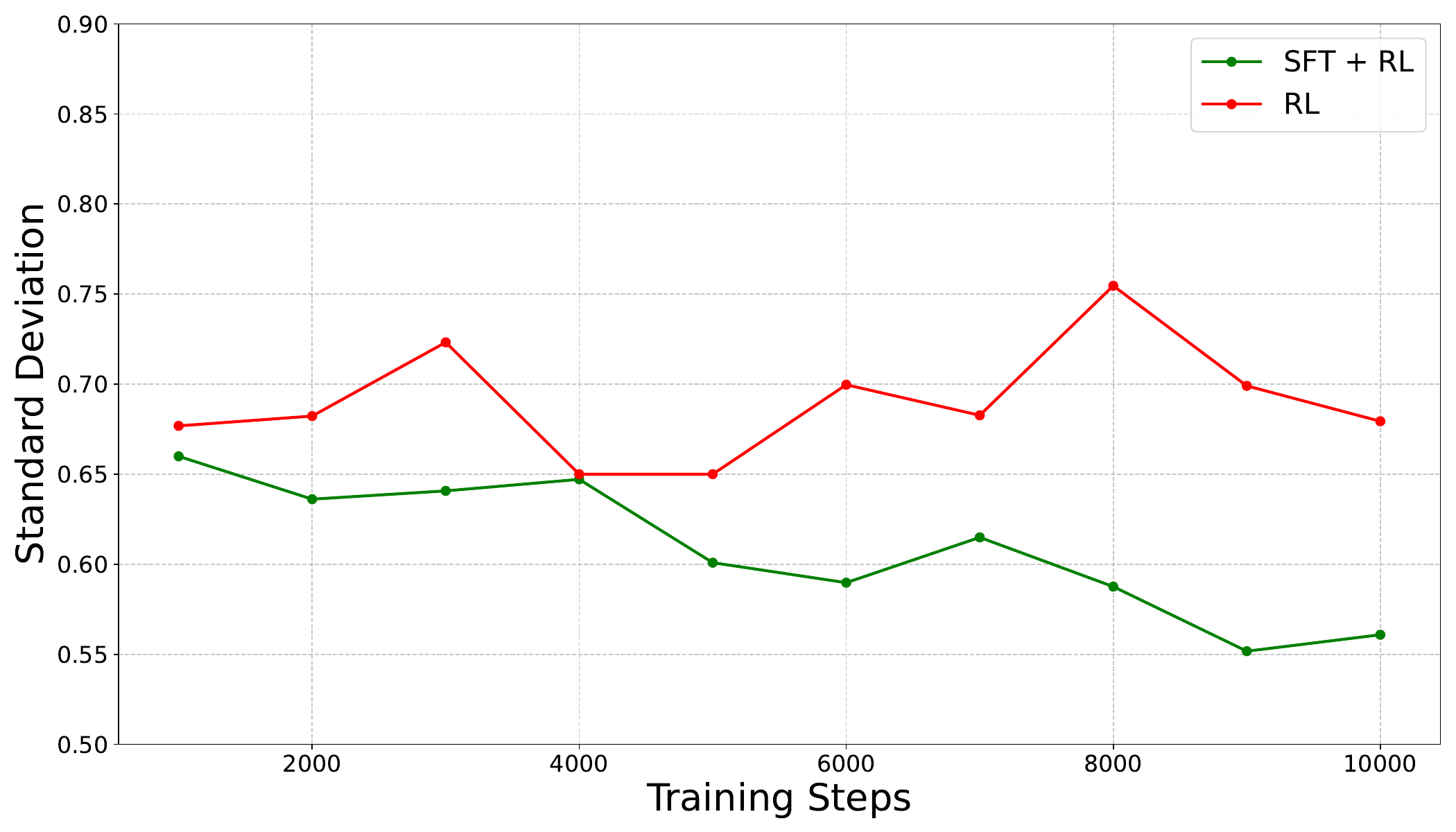}
       \caption{Standard Deviation of Rewards}
        \label{fig:std_reward}
    \end{subfigure}
 ~
    \begin{subfigure}[t]{0.45\textwidth}
        \centering
        \includegraphics[height=1.5in]{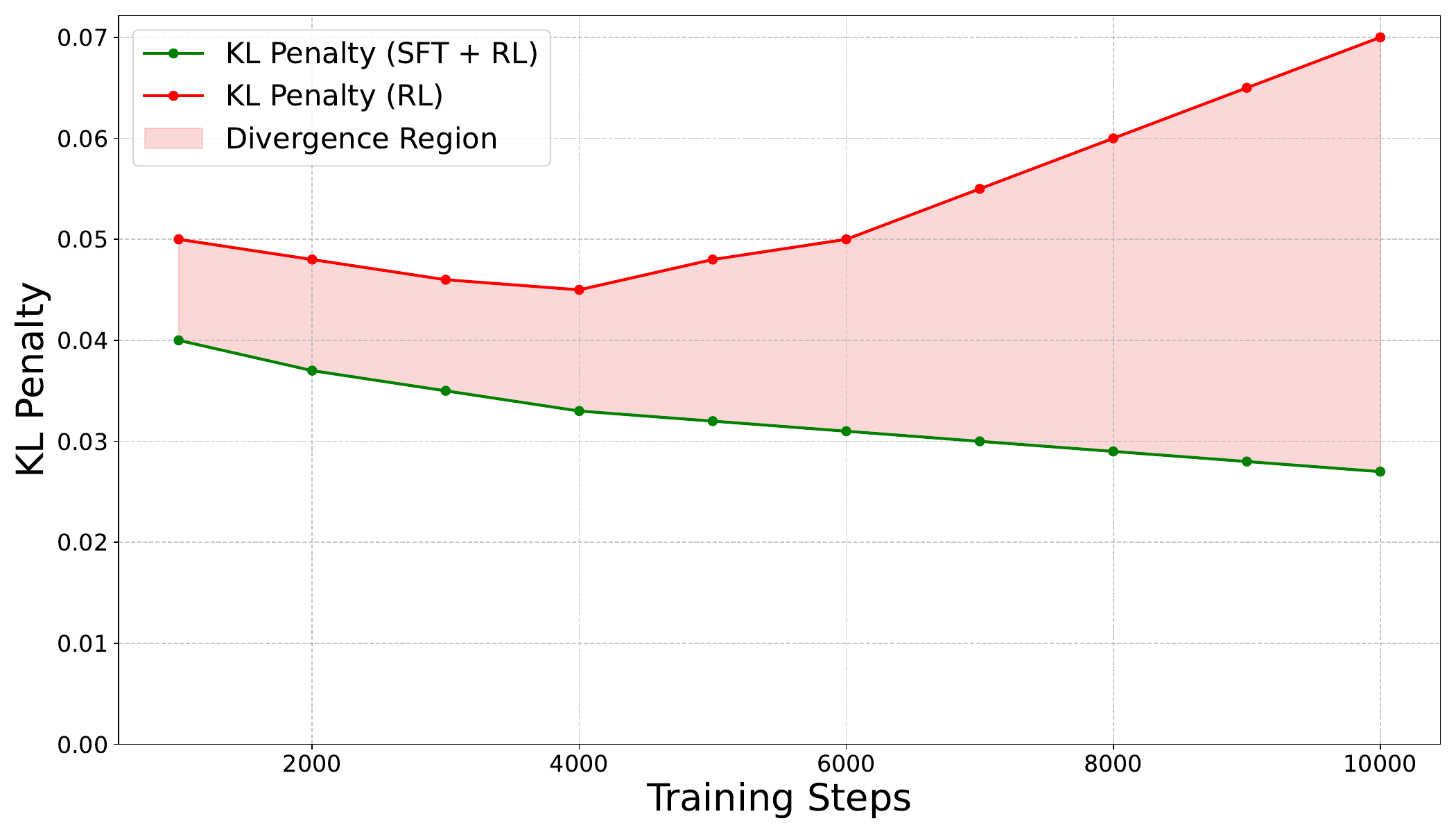}
        \caption{Negated KL Divergence Penalty}
        \label{fig:kl_penalty}
    \end{subfigure}
    \caption{RL training observations}
    \label{fig:rl_stats}
\end{figure*}

Figure~\ref{fig:rl_stats} illustrates the trends in the standard deviation of rewards and the KL-Divergence penalty across training steps. The initial decline in standard deviation, followed by stabilization, coupled with consistent KL-divergence penalties, suggests that our reward modeling strategy effectively aligns a fine-tuned language model for the extract method refactoring task. These results collectively highlight the efficacy of \rl{} in producing accurate extract method refactored code.

The combined approach addresses limitations associated with individual learning objectives, such as overfitting and vanishing gradient problems. While \textsc{sft} typically learns based on loss derived from labeled data, the integration with \textsc{rl} allows the model to benefit from a more comprehensive feedback system, including reward signals. This holistic approach contributes to the observed performance enhancements.

\begin{figure*}[ht!]
    \centering
    \includegraphics[width=\textwidth]{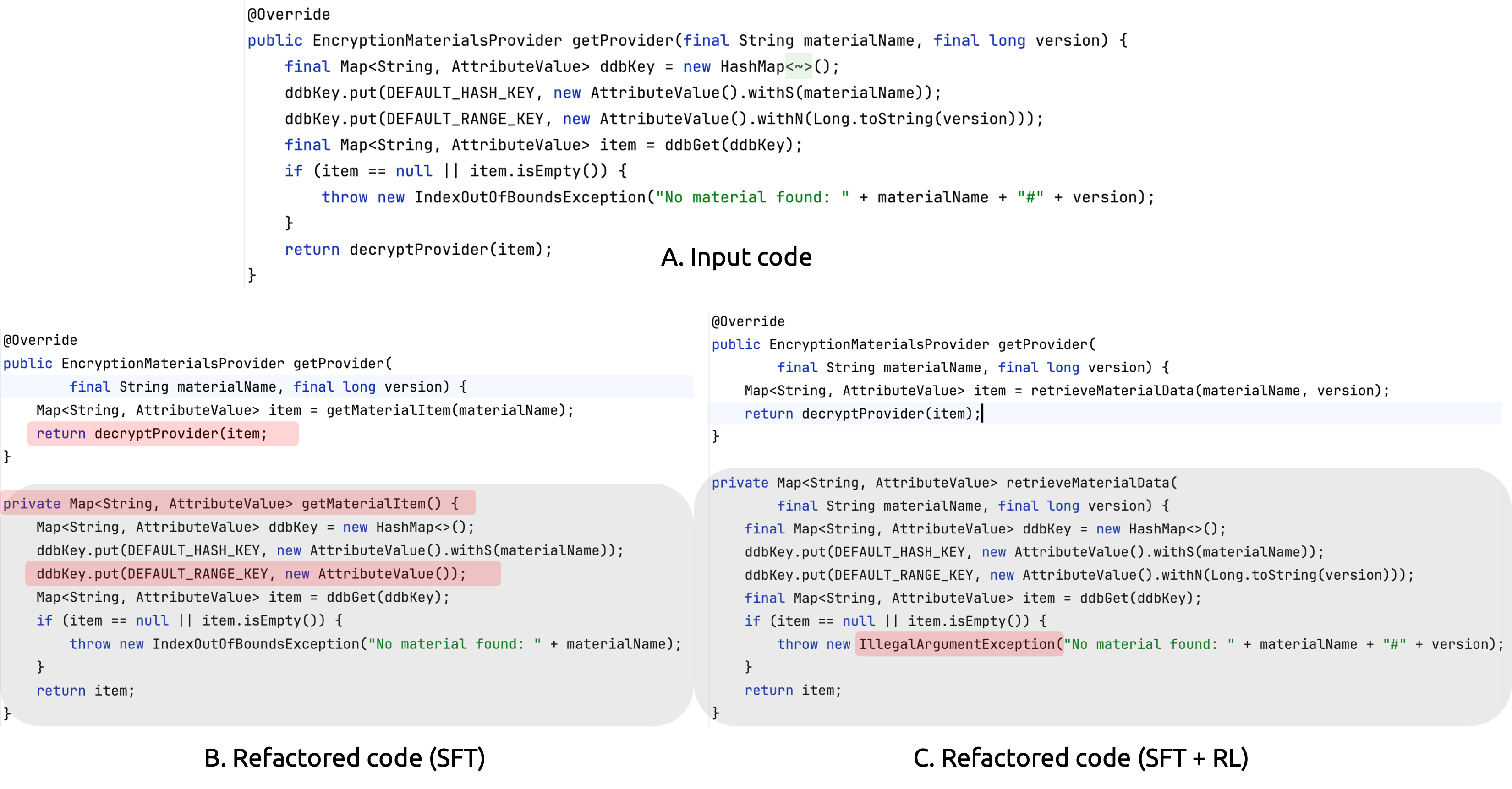}
    \caption{Extract method refactoring example generated using Supervised Fine-Tuning (SFT) and the combination of SFT and Reinforcement Learning (RL) techniques}
    \label{fig:example}
\end{figure*}

We present an illustrative example of a \exm{} refactoring in Figure~\ref{fig:example}, highlighting the differences of \sft{} and \rl{} techniques.
The original method belongs to \texttt{aws/aws-dynamodb-encryption\-java} repository as present in commit \texttt{ea43801}. 
Snippets \texttt{B} and \texttt{C} are generated by \sft{} model and combined \sft{} with \rl{} aligned models respectively. As we can see from the generated example, there are few syntactic errors (highlighted by red background color) present in the output generated by the fine-tuned only model. The combined \rl{} model seems to be more aligned to the ground truth. However, the generated code is not absolutely accurate because at line $10$ it throws an \texttt{IllegalArgumentException} instead of \texttt{IndexOutOfBoundException}. But this example strengthens our claims that the combined \sft{} model with \rl{} alignment enhances language model performance to generate more accurate extract method refactored code.

\begin{boxH}
    \textbf{RQ3 Summary:} Our results demonstrates that combining supervised fine-tuning and \rl{} objectives yields superior results in generating refactored code. This integrated approach outperforms individual methods, showing significant improvements in \codebleu{}, \bleu{}, and \rouge{} metrics, while mitigating common limitations associated with single-objective training.
\end{boxH}

\section{Discussions}

While statistical metrics provide valuable insights, they may not fully capture a model's ability to generate high-quality code. 
We can observe the phenomenon in Table \ref{tab:all_metrics} and Table \ref{tab:exresults}.
The evaluation metrics used in Table \ref{tab:all_metrics} do not show very drastic difference in the RQ1 and RQ3 results.
However, the qualitative results presented in Table \ref{tab:exresults} present very different narrative.
We observe that the models trained using both supervised fine-tuning and \rl{} techniques show significantly better results. Specifically, the number of test cases passed by the best model in RQ3 is 61\% more than that of the best model in RQ1.
This observation highlights the importance of qualitative evaluation in addition to traditional metrics-based evaluation.

One may wonder whether production-ready \llmsc{} for code, such as GitHub Copilot, can achieve better performance in automated extract method refactoring. Although this is beyond the scope of our study, it's reasonable to assume that such models could handle smaller, frequently performed refactorings. However, anecdotal evidence suggests these models may struggle with context-specific, complex, non-atomic refactorings. Further research is needed to examine the behavior of these models across various prompt settings and complexity levels.
\section{Related Work}

\subsection{Automated Refactoring}

Many studies have explored automated refactoring candidate identification using machine learning techniques. Typically, these studies use source code metrics or commit messages to train models. Aniche \etal{}~\cite{Aniche2020Effectiveness} predict $20$ kinds of refactorings at method, class, or variable levels using code, process, and ownership metrics, with Random Forest performing best among six algorithms. Gerling~\cite{Gerling2020Machine} extended this work by improving the data collection process to create a high-quality, large-scale refactoring dataset. Van Der Leij \etal{}~\cite{van2021data} analyze five machine learning models to predict Extract Method refactoring, comparing results with industry experts. Using $61$ code metrics, they also found Random Forest to be the best performing model. Kumar \etal{}~\cite{kumar2019method} studied method-level refactoring prediction, analyzing 10 machine learning classifiers. Sagar \etal{}~\cite{sagar2021comparing} approaches refactoring candidate prediction as a multi-class classification problem, using both source code quality metrics and commit messages as features to predict six method-level refactorings. They compare text-based and source code-based models. Kurbatova \etal{}~\cite{Kurbatova2020Recommendation} employ code embeddings generated from Code2Vec~\cite{alon2019code2vec} to train their model for Move Method refactoring prediction.

In the domain of automated code refactoring, researchers have developed a variety of specialized tools and approaches. CeDAR~\cite{tairas2012cedar}, an Eclipse plugin, focuses on identifying and eliminating duplicate code. JDeodorant~\cite{JDeodrant, mazinanian2016jdeodorant} detects code smells and proposes refactoring strategies. Fokaefs~\etal{}\cite{fokaefs2012identification} extended JDeodorant's capabilities to prioritize and implement class extraction refactorings. SOMOMOTO\cite{zanetti2014automated} facilitates move method refactoring and code modularization. While these rule-based methods have made significant contributions, they face limitations in capturing semantic information during refactoring. Moreover, they often require manual intervention from developers to identify and select code blocks for refactoring. To address these challenges, recent research has explored the application of deep learning and large language models (LLMs) for automated code refactoring.

Szalontai~\etal{}~\cite{szalontai2023deep} developed a deep learning method for refactoring source code, initially designed for the Erlang programming language. Their approach comprises a localizer and a refactoring component, enabling the identification and transformation of non-idiomatic code patterns into idiomatic counterparts. Tufano~\etal{}~\cite{tufano2019learning} conducted a quantitative investigation into the potential of Neural Machine Translation (NMT) models for automatically applying code changes implemented during pull requests. Their approach leverages NMT to translate code components from their pre-pull request state to their post-pull request state, effectively simulating developer-implemented changes. To facilitate the rename refactoring process and reduce cognitive load on developers, Liu~\etal{}~\cite{liu2023refbert} proposed RefBERT, a two-stage pre-trained framework based on the BERT architecture. RefBERT is designed to automatically suggest meaningful variable names, addressing a challenging aspect of code refactoring.

Current automated refactoring tools lack semantic understanding and require manual intervention. To address this, we propose a hybrid approach combining supervised fine-tuning with reinforcement learning, enhancing the accuracy and completeness of extract method refactoring. This is the first study to apply deep reinforcement learning for this task, advancing automated refactoring tools.

\subsection{Reinforcement learning in software engineering}

Sequence modeling has emerged as a fundamental paradigm for addressing a wide array of software engineering challenges. In recent years, researchers have explored the application of deep reinforcement learning (DRL) techniques to mitigate exposure bias in supervised fine-tuned models for sequence generation tasks~\cite{ranzato2016sequenceleveltrainingrecurrent, keneshloo2019deepreinforcementlearningsequence}. Notably, Ranzato~\etal{}~\cite{ranzato2016sequenceleveltrainingrecurrent} pioneered the use of established metrics such as BLEU and ROUGE as reward signals in DRL algorithms to optimize network parameters in machine translation, effectively addressing exposure bias. The intersection of DRL and sequence modeling has led to innovative frameworks, such as the one proposed by Chen~\etal{}~\cite{chen2021decisiontransformerreinforcementlearning}, which reconceptualizes reinforcement learning problems as sequence modeling tasks. This approach has paved the way for novel applications in various domains.

In the realm of software engineering, DRL methods have gained traction, particularly in code completion and summarization tasks. Wang~\etal{}\cite{wang2022compilable} leveraged compiler feedback as a reward signal to enhance the quality of language model-generated code. Le\etal{}\cite{le2022coderl} introduced CodeRL, a framework that integrates RL with unit test signals to fine-tune program synthesis models. Shojaee~\etal{}~\cite{shojaee2023execution} conducted comprehensive research, proposing a framework for fine-tuning code language models using DRL and execution signals as rewards. Recent advancements in this field include IRCOCO by Li~\etal{}\cite{li2024ircoco}, which employs immediate rewards to fine-tune language models for code completion tasks. Wang\etal{}\cite{wang2024rlcoder} developed RLCoder, combining DRL with Retrieval-Augmented Generation (RAG) pipelines for repository-level code completion. Furthermore, Nichols\etal{}~\cite{nichols2024performance} demonstrated the potential of DRL in generating efficient parallel code, expanding the application of these techniques to performance optimization.

To our knowledge, \llm{}s have not been specifically trained or aligned for extract method refactoring. Our approach, which combines supervised fine-tuning with PPO alignment, is a first in this domain. This novel methodology produces accurate refactored methods, marking a significant advancement in the field.
\section{Threats to Validity}

\textit{\textbf{Internal validity:}} Internal validity concerns relate to the reliability of conclusions drawn from our experimental results. To enhance the trustworthiness of our findings, we implemented several measures. Firstly, we addressed the potential confounding effect of varying hyperparameters by utilizing consistent settings across all models, based on the optimal configurations identified in prior research by Li~\etal{}~\cite{li2024ircoco}. Additionally, we employed identical data splits for training and testing across all models, ensuring equitable learning opportunities and evaluation conditions. These methodological decisions mitigate the risk of spurious results attributable to inconsistent experimental conditions, thereby strengthening the validity of our conclusions regarding the efficacy of deep reinforcement learning in generating refactored code methods.

\textit{\textbf{External validity:}} External validity concerns in our study pertain to the generalizability of our findings beyond the Java context. Despite this focus, we argue that our methodology is highly transferable. Our data collection technique is language-agnostic, applicable to any refactoring scenario. The general-purpose models we employed, trained on vast code corpora, are adaptable to various programming languages. While these factors suggest broad applicability, further research across multiple languages and environments would be necessary to conclusively establish the universal validity of our approach.
\section{Conclusions}

In this study, we introduce a novel approach that integrates traditional fine-tuning with reinforcement learning alignment to automatically generate extract method refactorings for Java code.
To evaluate the generated code, we not only rely on traditional metrics such as \bleu{} and \rouge{}
but also construct a detailed qualitative evaluation mechanism to check the syntactic and semantic correctness.
Experimental results demonstrate that our approach significantly improves the performance of large language models in code refactoring compared to supervised fine-tuning, as evidenced by quantitative evaluation metrics and qualitative measures.

Our future research will focus on expanding the scope of our approach to encompass various types of refactorings, different programming languages, and industry-based codebases. We also plan to increase the size of our dataset, especially the qualitative evaluation set, for more comprehensive evaluations. We also plan to explore the use of automated test suite generation tools such as EvoSuite~\cite{evosuite} to expand our sample size. Furthermore, we intend to investigate alternative reinforcement learning algorithms and reward strategies to further enhance the performance and effectiveness of our automated code refactoring approach.

\textbf{Data Availability}: Our replication package including source code and data is available online at \href{https://anonymous.4open.science/r/extract-method-generation-2C9B}{https://anonymous.4open.science/r/extract-method-generation-2C9B}.

\bibliographystyle{ACM-Reference-Format}
\bibliography{references}

\appendix

\end{document}